\pdfoutput=1
\RequirePackage{color}
\documentclass{JINST}
\usepackage{xspace}
\usepackage{multirow}
\usepackage{lineno}

\def\mm {\ensuremath{\mbox{\,mm}}\xspace}
\newcommand{\gev}{\ensuremath{\mathrm{\,Ge\kern -0.1em V}}\xspace}
\newcommand{\tev}{\ensuremath{\mathrm{\,Te\kern -0.1em V}}\xspace}
\def\pt         {\mbox{$p_{\rm T}$}\xspace}

\title{Mapping the material in the LHCb vertex locator using secondary hadronic interactions}

\author{
M.~Alexander$^{1}$,
W.~Barter$^{2}$,
A.~Bay$^{3}$,
L.J.~Bel$^{4}$,
M.~van~Beuzekom$^{4}$,
G.~Bogdanova$^{5}$, %
S.~Borghi$^{2}$,
T.J.V.~Bowcock$^{6}$,
E.~Buchanan$^{7}$,
J.~Buytaert$^{8}$,
K.~Carvalho~Akiba$^{9}$,
S.~Chen$^{10}$,
V.~Coco$^{8}$, %
P.~Collins$^{8}$,
A.~Crocombe$^{11}$,
F.~Da~Cunha~Marinho$^{9}$,
E.~Dall'Occo$^{4}$,
S.~De~Capua$^{2}$,
C.T.~Dean$^{1}$,
F.~Dettori$^{6}$,
D.~Dossett$^{11,a}$,
K.~Dreimanis$^{6}$,
G.~Dujany$^{12}$,
L.~Eklund$^{1}$,
T.~Evans$^{13}$,
M.~Ferro-Luzzi$^{8}$,
M.~Gersabeck$^{2}$,
T.~Gershon$^{11}$,
T.~Hadavizadeh$^{13}$,
J.~Harrison$^{2}$, %
K.~Hennessy$^{6}$,
W.~Hulsbergen$^{4}$,
D.~Hutchcroft$^{6}$,
P.~Ilten$^{14,b}$\footnote{corresponding authors: philten@cern.ch, mwill@mit.edu},
E.~Jans$^{4}$,
M.~John$^{13}$,
P.~Kopciewicz$^{15}$, %
P.~Koppenburg$^{4}$,
G.~Lafferty$^{2}$,
T.~Latham$^{11}$,
A.~Leflat$^{5,8}$,
M.W.~Majewski$^{15}$,
R.~McNulty$^{16}$,
J.~Mylroie-Smith$^{6}$, %
A.~Oblakowska-Mucha$^{15}$,
C.~Parkes$^{2}$,
A.~Pearce$^{8}$,
A.~Poluektov$^{11}$,
A.~Pritchard$^{6}$, %
W.~Qian$^{11}$,
S.~Redford$^{13}$, %
S.~Richards$^{7}$,
K.~Rinnert$^{6}$,
E.~Rodrigues$^{17}$,
G.~Sarpis$^{2}$,
M.~Schiller$^{1}$,
H.~Schindler$^{8}$,
M.~Smith$^{2}$,
N.A.~Smith$^{6}$, %
T.~Szumlak$^{15}$,
J.J.~Velthuis$^{7}$,
V.~Volkov$^{5}$,
C.~Wallace$^{11}$,
H.M.~Wark$^{6}$,
A.~Webber$^{2}$, %
M.R.J.~Williams$^{2}$,
M.~Williams$^{14*}$. \\ \vspace{-0.05in}

\begin{footnotesize}
$ ^{1}$School of Physics and Astronomy, University of Glasgow, Glasgow, United Kingdom\\
$ ^{2}$School of Physics and Astronomy, University of Manchester, Manchester, United Kingdom\\
$ ^{3}$Institute of Physics, Ecole Polytechnique  F{\'e}d{\'e}rale de Lausanne (EPFL), Lausanne, Switzerland\\
$ ^{4}$Nikhef National Institute for Subatomic Physics, Amsterdam, The Netherlands\\
$ ^{5}$Institute of Nuclear Physics, Moscow State University (SINP MSU), Moscow, Russia\\
$ ^{6}$Oliver Lodge Laboratory, University of Liverpool, Liverpool, United Kingdom\\
$ ^{7}$H.H. Wills Physics Laboratory, University of Bristol, Bristol, United Kingdom\\
$ ^{8}$European Organization for Nuclear Research (CERN), Geneva, Switzerland\\
$ ^{9}$Universidade Federal do Rio de Janeiro (UFRJ), Rio de Janeiro, Brazil\\
$ ^{10}$Sezione INFN di Cagliari, Cagliari, Italy\\
$ ^{11}$Department of Physics, University of Warwick, Coventry, United Kingdom\\
$ ^{12}$LPNHE, Universit{\'e} Pierre et Marie Curie, Universit{\'e} Paris Diderot, CNRS/IN2P3, Paris, France\\
$ ^{13}$Department of Physics, University of Oxford, Oxford, United Kingdom\\
$ ^{14}$Massachusetts Institute of Technology, Cambridge, MA, United States\\
$ ^{15}$AGH - University of Science and Technology, Faculty of Physics and Applied Computer Science, Krak{\'o}w, Poland\\
$ ^{16}$School of Physics, University College Dublin, Dublin, Ireland\\
$ ^{17}$University of Cincinnati, Cincinnati, OH, United States\\
$ ^{a}$Current address: School of Physics, The University of Melbourne, VIC 3010, Australia \\
$ ^{b}$Current address: School of Physics and Astronomy, University of Birmingham, Birmingham, United Kingdom
\end{footnotesize}
}

\abstract{
Precise knowledge of the location of the material in the LHCb vertex locator (VELO) is essential to reducing background in searches for long-lived exotic particles, and in identifying jets that originate from beauty and charm quarks.
Secondary interactions of hadrons produced in beam-gas collisions are used to map the location of material in the VELO.
Using this material map, along with properties of a reconstructed secondary vertex and its constituent tracks, a $p$-value can be assigned to the hypothesis that the secondary vertex originates from a material interaction.
A validation of this procedure is presented using photon conversions to dimuons.
}

\begin{document}

\section{Introduction}

Precise knowledge of the location of the material in the LHCb vertex locator (VELO)~\cite{ref:velo} is essential to reducing background in searches for long-lived exotic particles (see, {\em e.g.}, Refs.~\cite{ref:atomm,ref:atoee}), and in identifying jets that originate from beauty and charm quarks~\cite{ref:bcjets}.
The VELO aperture is smaller than required by the LHC beams during injection;
therefore, the VELO consists of two retractable halves that close about the interaction region after the LHC beams are stable, resulting in fill-by-fill movement of the VELO material.
The location of the $pp$-collision region changes by $\mathcal{O}(0.1\mm)$ from year to year, and changed by $\approx 0.5\mm$ between the first and second LHC runs.
As part of the LHCb data calibration process, a software-based alignment procedure is used to precisely determine the location of each VELO module, and in addition, to account for the fill-by-fill movement of each VELO half~\cite{ref:veloalign}.
This alignment procedure, however, can only precisely determine the location of active sensor elements.
An alternative approach is required to fully map the VELO material.

This article presents a study of the VELO material based on secondary hadronic interactions, an approach previously used by LHCb to determine the VELO aperture~\cite{ref:veloap} and widely used by other experiments to study detector material (see, {\em e.g.}, Ref.~\cite{ref:atlasmat}).
The typical resolution on the reconstructed vertex position for a secondary hadronic interaction is about 0.04\mm in $r$ and 0.4\mm in $z$,\footnote{LHCb defines its coordinate system as follows: $\hat{z}$ is along the beam line, where positive $z$ denotes the direction from the $pp$-interaction point into the LHCb detector; $\hat{y}$ is vertical upwards; and $\hat{x}$ is horizontal and defined such that the coordinate system is right handed. Polar coordinates, $r$ and $\phi$, are also used in the $xy$ plane.}
 which enables the construction of a high-precision map of the location of material in the VELO.
Using this material map, along with properties of a reconstructed secondary vertex (SV) and its constituent tracks, a $p$-value can be assigned to the hypothesis that the SV originates from a material interaction.
This approach was recently used to veto photon conversions to $\mu^+\mu^-$ in a search for dark photons at LHCb~\cite{ref:lhcbatomm}.\footnote{Many other experiments have employed alternative strategies for rejecting material-induced backgrounds in searches for long-lived particles, {\em e.g.}, see Ref.~\cite{ref:atlasll}.}

This analysis uses secondary interactions of hadrons produced in beam-gas collisions collected during special run periods where helium or neon gas was injected into the beam-crossing region~\cite{ref:smog2}.
In such events, the LHC proton beams collide with the gas molecules producing primary hadrons, which subsequently scatter off the VELO material producing secondary hadrons that are detected by LHCb.
Material interactions occur along the entire length of the VELO in beam-gas events, rather than just near the $pp$-interaction region.
A modified tracking configuration that makes no assumptions about the particle origin points along $z$
is used to reconstruct the particles produced in the secondary interactions.
This article is structured as follows:
the VELO detector, beam-gas data sets, and track and vertex reconstruction algorithms are described in Sec.~\ref{sec:data};
in Sec.~\ref{sec:map}, the material map is presented;
the procedure for obtaining material interaction $p$-values is discussed in Sec.~\ref{sec:prob};
and Sec.~\ref{sec:sum} summarizes.

\section{Detector and Data Sets}
\label{sec:data}

\begin{figure}[t]
  \centering
  \includegraphics[height=0.25\textheight]{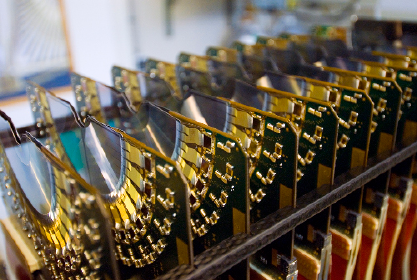}
  \includegraphics[height=0.25\textheight]{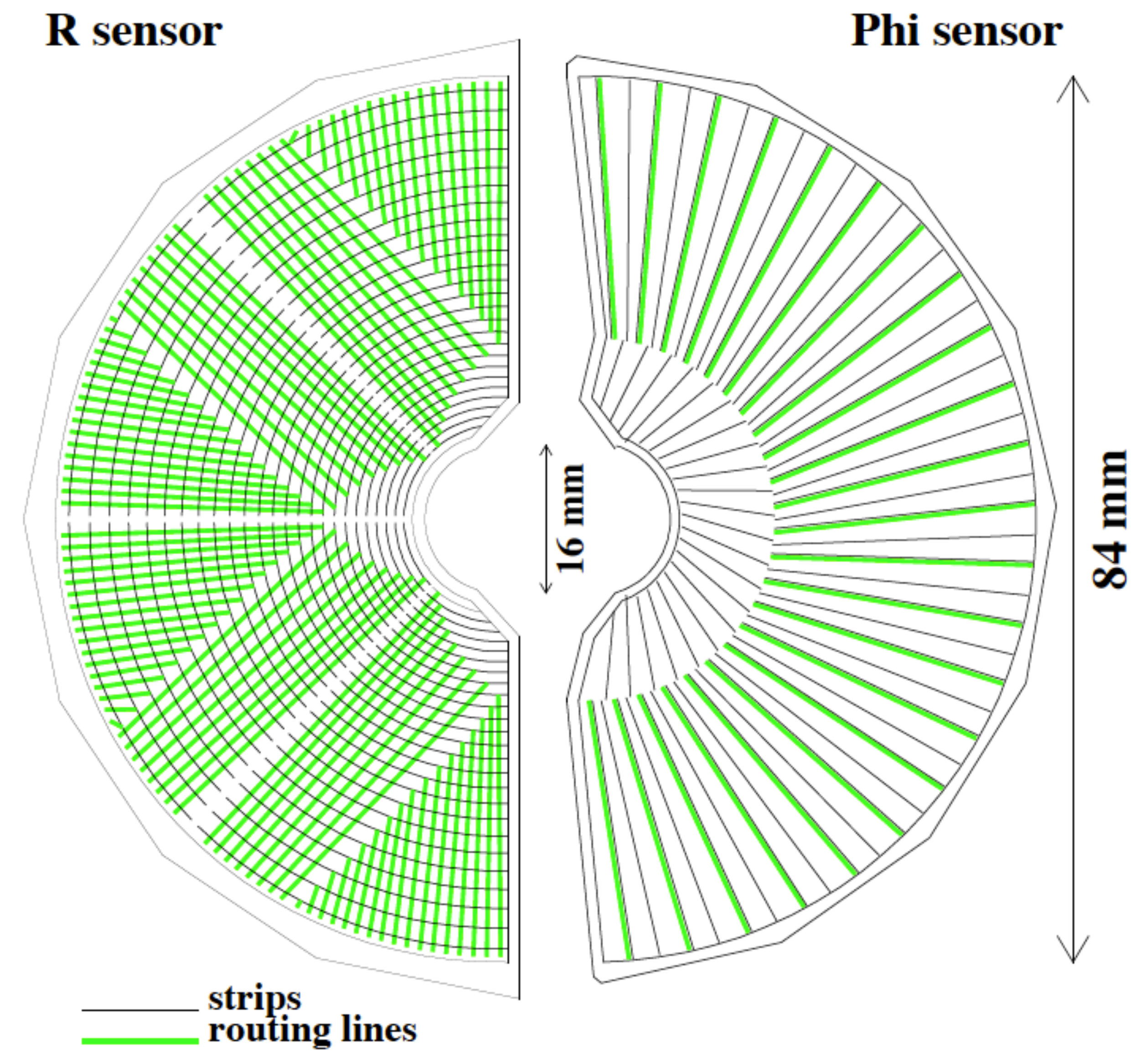}\\
  \includegraphics[width=0.99\textwidth]{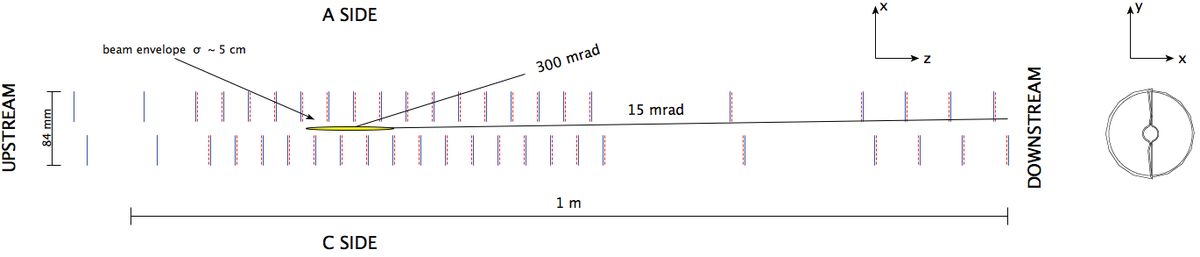}
  \caption{
  From Ref.~\cite{ref:velo}:
  (top left) a photograph of one side of the VELO, taken during assembly, showing the silicon sensors and readout hybrids;
  (top right) a schematic of both an $r$ and $\phi$ sensor, showing the sensor strips and routing lines;
  and
  (bottom) schematics showing the cross section of the $xz$ plane at $y=0$, where the $r(\phi)$ sensors are shown with solid blue (dashed red) lines,
  and an $xy$ view of overlapping sensors in the closed position.
  {\em N.b.}, the modules at positive (negative) $x$ are known as the left or A-side (right or C-side).
  }
  \label{fig:velo}
\end{figure}

The LHCb detector is a single-arm spectrometer covering the forward pseudorapidity region of ${2 < \eta < 5}$~\cite{ref:lhcb}.
The detector, which was built to study the decays of hadrons containing $b$ and $c$ quarks, includes a high-precision charged-particle tracking system, two ring-imaging Cherenkov detectors, electromagnetic and hadronic calorimeters, and a system of muon chambers.
The LHCb collaboration mostly collected $pp$-collision data at center-of-mass energies of 7 and 8\tev in Run~1 and  at 13\tev in Run~2; however, special running periods at alternative energies, using heavy-ion beams
and gaseous targets
have also been undertaken.

The VELO is a silicon-microstrip detector that surrounds the $pp$-interaction region and provides excellent vertex resolution (see Fig.~\ref{fig:velo}).
During physics data taking, the VELO sensors are moved to within 7\mm of the beam, with the closest active regions only about 8\mm in the transverse plane from  the $pp$ collisions.
This enables achieving a charged particle impact parameter resolution of 0.035\mm for a transverse momentum of $\pt \approx 1\gev$, and as low as 0.012\mm for high-momentum particles.
A detailed description of the VELO performance is provided in Ref.~\cite{ref:velo}.

Each VELO half contains 42 silicon-microstrip sensors roughly semi-circular in shape with an outer radius of 42\mm, an excised inner semi-circle of radius 7\mm, and a thickness (in $z$) of 0.3\mm.
These sensors are placed into 21 {\em standard} modules in $r$--$\phi$ pairs, where each $r$ sensor provides a radial-coordinate measurement, each $\phi$ sensor provides an azimuthal-coordinate measurement, and the module location determines the $z$ coordinate.
Each half also has two additional modules, referred to as the pile-up system, that only contain an $r$ sensor and are located in the most upstream positions of the VELO.
There is a slight overlap between the two VELO halves to ensure full angular coverage and to assist in calibration of the detector.
Each  half is contained in an {\em RF-box}, which provides an independent vacuum from the LHC machine vacuum.
The beam-facing surface of the RF-boxes is the {\em RF-foil}, a 0.3\mm thick AlMg$_3$ sheet that is corrugated around the modules to minimize the material traversed by charged particles.
The RF-box and RF-foil shield the VELO sensors against RF pickup from the LHC beams, prevent impedance disruptions of the LHC beams, and protect the LHC machine vacuum.

\begin{figure}[t]
  \centering
  \includegraphics[width=0.49\textwidth]{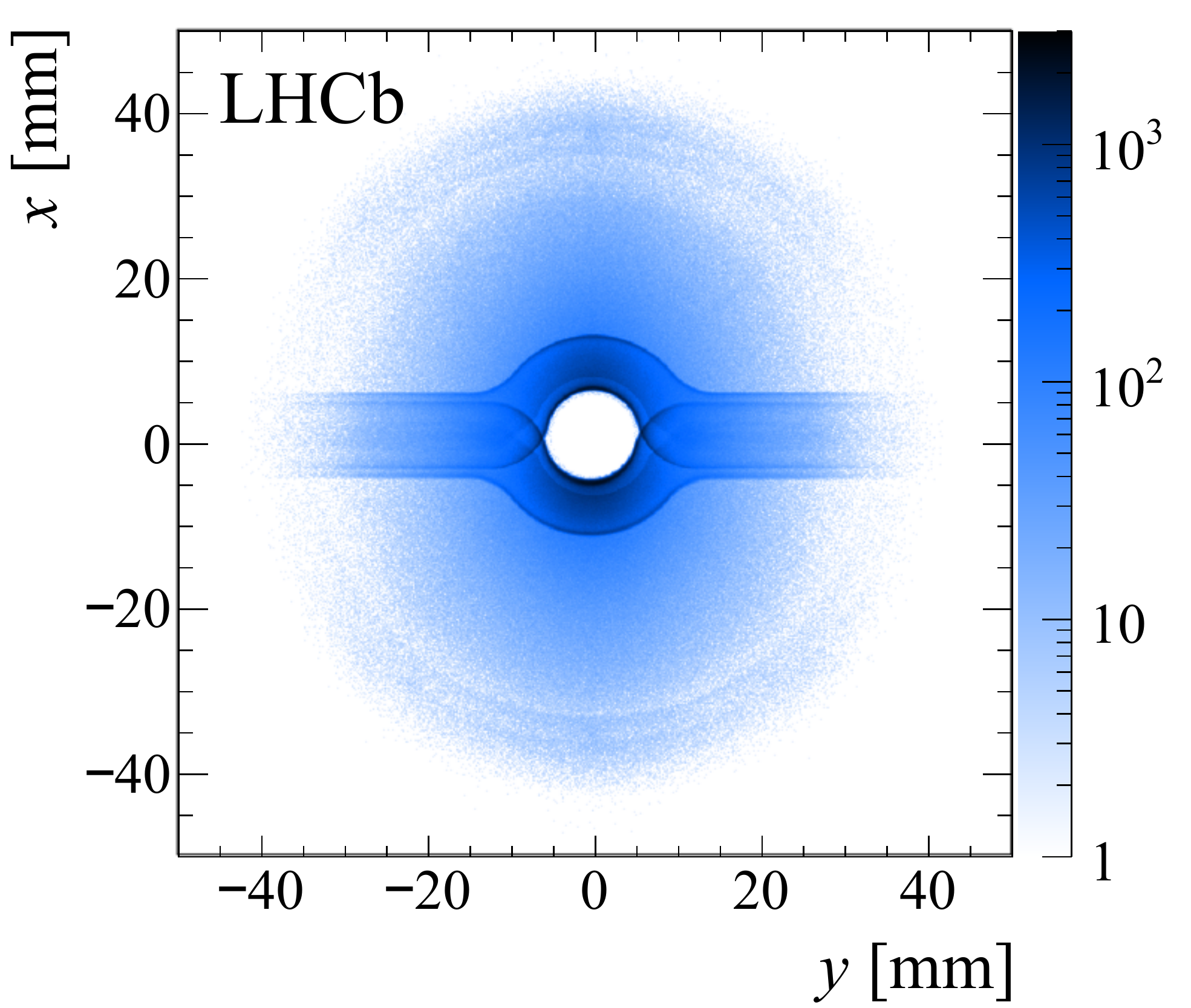}
  \includegraphics[width=0.49\textwidth]{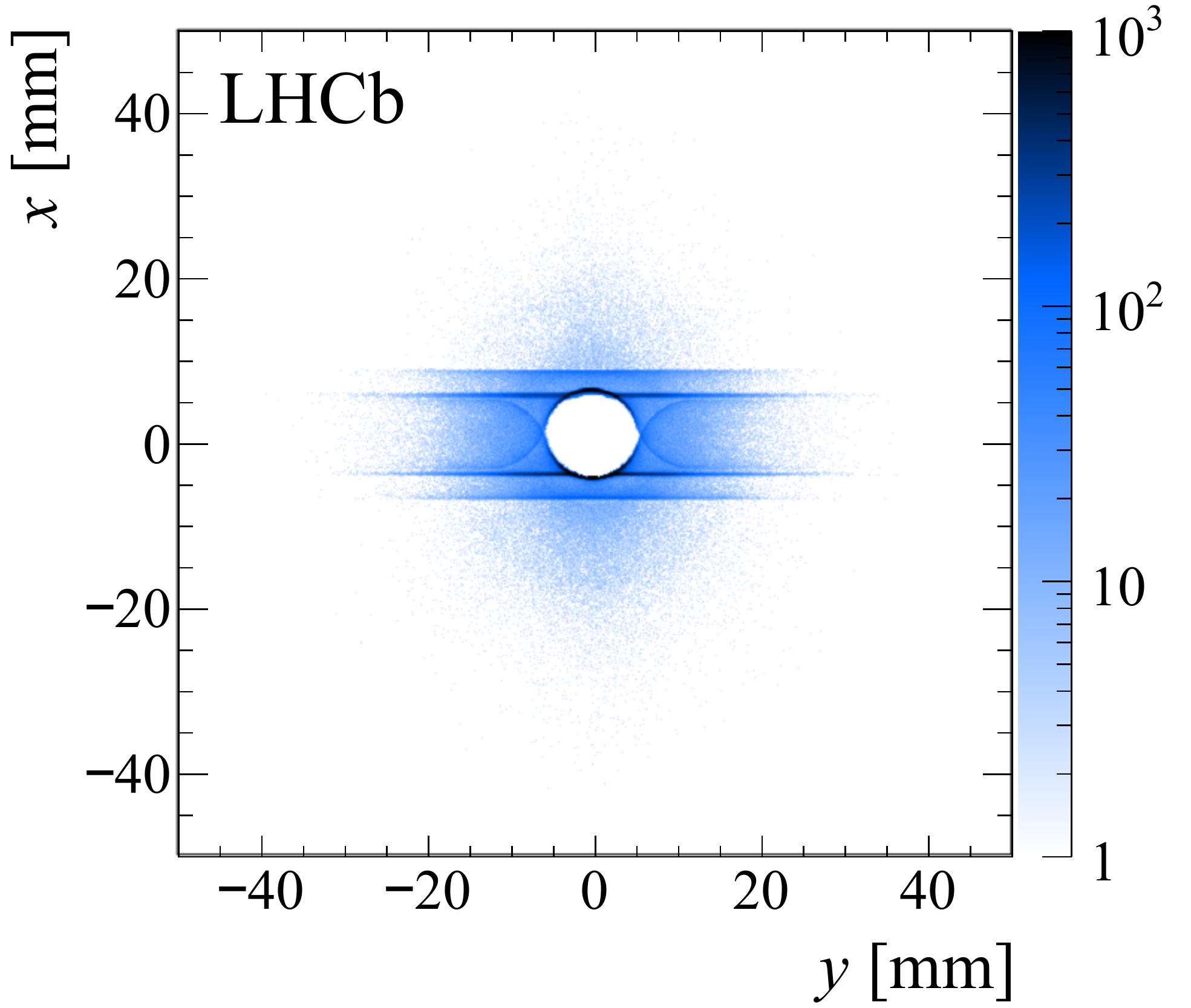}
  \caption{
  Reconstructed SVs in the Run~2 data sample showing the $xy$ plane integrated over $z$ within the region of the VELO that contains sensor modules.
  The left (right) panel shows the central (forward) VELO region.
  The bins are $0.1\mm\times0.1\mm$ in size.
  }
  \label{fig:svs1}
\end{figure}

\begin{figure}[t]
  \centering
  \includegraphics[width=0.99\textwidth]{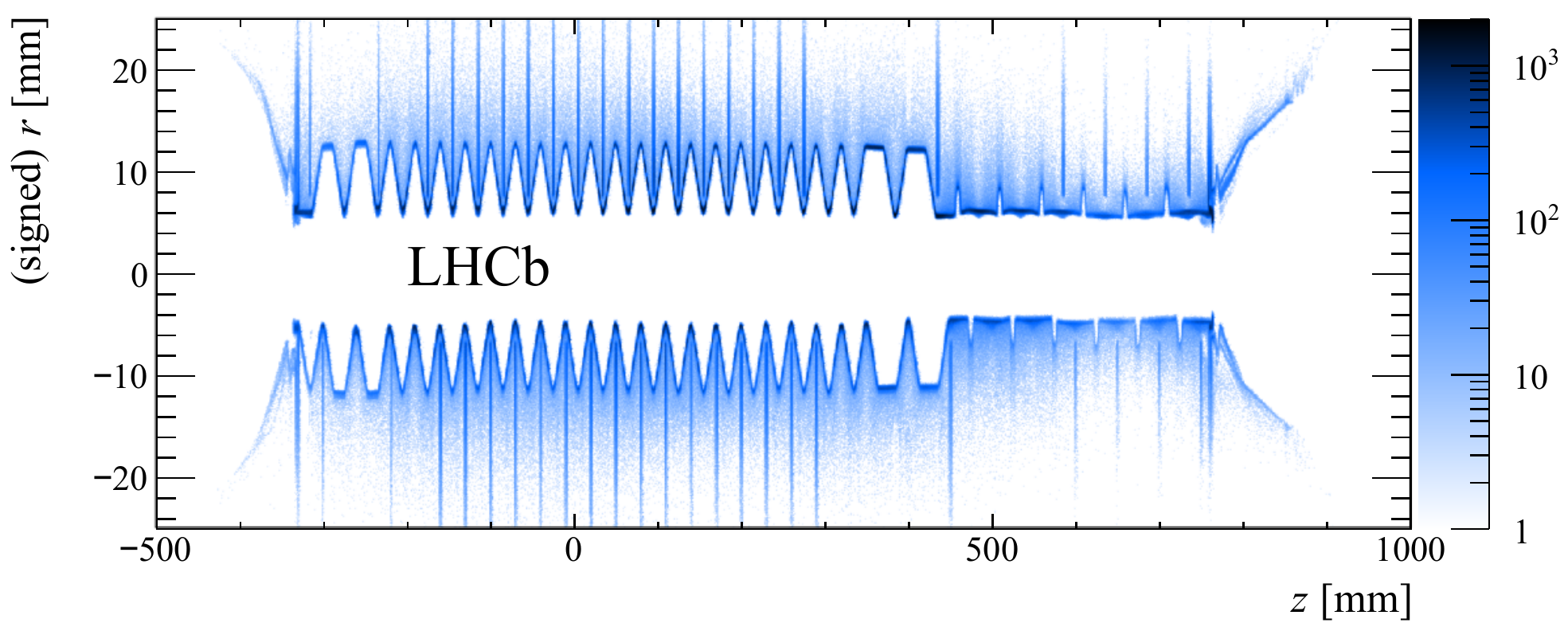}
  \caption{
  Reconstructed SVs in the Run~1 data sample showing the $zr$ plane integrated over $\phi$, where a positive (negative) $r$ value denotes that the SV is closest to material in the right (left) half of the VELO.
  The bins are $0.1\mm\times1\mm$ in size.
  {\em N.b.}, the inner-most RF-foil region is nearly semi-circular in the $xy$ plane, which results in sharp edges at smaller $r$ values; however, at large $|y|$ values, the RF-foil is flat producing SVs at larger values of $r$ which can easily be mistaken as background in the $zr$ projection shown here.
  }
  \label{fig:svs2}
\end{figure}

A beam-gas imaging system was proposed in Ref.~\cite{ref:smog1}, and developed and commissioned during Run~1~\cite{ref:smog2} to enable making high-precision luminosity measurements~\cite{ref:lumi}.
This system has since been repurposed to allow LHCb to collect data as a fixed-gaseous-target experiment.
The analysis presented here uses secondary interactions of hadrons produced in beam-gas collisions collected during special run periods where helium or neon gas was injected into the beam-crossing region.
Several data samples are used from different running periods:
data collected during $pp$ running in 2011 and 2012 (Run~1) meant for luminosity studies, with beam energies of 3.5 and 4.0\tev, respectively;
and
data taken during a dedicated proton-helium run in 2016 (Run~2) with a beam energy of 4.0\tev.
Only one LHC beam has a nominally filled bunch slot in all events used in this study.
The data sets were collected using minimum bias triggers.

Since the particles produced in secondary interactions in beam-gas events do not necessarily originate from near the interaction point or the beam line,
the tracks used in this analysis are reconstructed using a modified tracking configuration that makes no assumptions about the origins along $z$ of the particles.
All reconstructed tracks are required to be of good quality and to have hits in at least 3 $r$--$\phi$ sensor pairs.
The SVs are reconstructed from 3 or more tracks and are required to be of good quality.
Futhermore, the SVs are required to be inconsistent with originating from a primary beam-gas collision, and only events with exactly one SV are used.
In total, the Run~1 and Run~2 data samples contain 14M and 38M SVs, respectively.
Figures~\ref{fig:svs1} and \ref{fig:svs2} present some displays of the reconstructed SV locations.

\section{Material Maps}
\label{sec:map}

The VELO closes around the beams at the beginning of each fill with a precision of $\mathcal{O}(0.01\mm)$~\cite{ref:velo}.
As stated above, the location of the $pp$-collision region (beam spot) changes by $\mathcal{O}(0.1\mm)$ from year to year, and changed by $\approx 0.5\mm$ between Run~1 and Run~2.
Separate VELO material maps are constructed for Run~1 and Run~2 to also allow for shifts of the VELO module locations relative to each other or relative to the RF-foil; however,
it is found that the VELO material is consistent with having only moved globally by the amount expected due to the change in the beam spot location, and only a single map is presented below.
This map must be adjusted for the beam-spot location of each data-taking period when used in an LHCb analysis.

The $z$ positions of the sensors are determined by fitting the observed SV $z$ distributions near each module location.
In these fits, the SVs are required to have $r > 7\mm$ and satisfy $x > -1.5\mm$ $(x < 1.5\mm)$ for the left (right) VELO half.
These requirements highly suppress contributions from material interactions in the RF-foil and from beam-induced backgrounds.
The fits estimate the center-of-mass of each sensor in $z$ which is stored as part of the material map.
The sensors are 0.3\mm thick and a material interaction can occur anywhere along the path of a particle traversing a sensor.
This is accounted for in the material-interaction $p$-values by assigning an uncertainty to the true origin point in the sensor.
Furthermore, the sensors are tilted, {\em i.e.}\ not aligned at a single $z$ value.
This effect is small compared to the SV resolution;\footnote{The resolution on the $z$ positions of the SVs used in this study is $\approx 0.4\mm$ on average. The difference in the $z$ position of a sensor at different values of $r$ due to tilts is typically $\lesssim 0.05\mm$ with the largest difference in $z$ being $\approx 0.15\mm$, determined by the LHCb software-based alignment procedure which does account for tilt. }
 therefore, it is accounted for using an additional uncertainty when determining the $p$-values, rather than being individually determined for each sensor.
The precision of these SV-based fits in determining the $z$ locations of the sensors is estimated to be $0.03\mm$ by comparing the distance between the $r$-sensor and $\phi$-sensor in each $r$--$\phi$ pair to the values determined by  metrology during VELO construction.

Since the manufacturing tolerance of the sensor wafers is only 0.05\mm, the nominal wafer shapes in the transverse plane are used for the sensors; {\em i.e.}\ the shapes of the modules are not varied in this study, only their locations.
The reference point used to define the $x$ and $y$ positions of each sensor, which is taken to be the center of the module semi-circles nominally at $x_{\rm beam}$ and $y_{\rm beam}$, is fitted to the observed $xy$ positions using SVs near each sensor in $z$.
Only SVs that are consistent with originating from an interaction in a module are used in these fits.
The $xy$ location of each module reference point is varied in order to maximize the vertex density that overlaps the module location.\footnote{Since the module shapes are fixed, moving the location of the reference point results in a translation in $xy$ of the entire module.}
The precision of these fits in determining the $xy$ locations of the sensors is estimated to be $0.03\mm$ (in each direction) by comparing the SV-based results to the high-precision software-based alignment locations.

\begin{figure}[t]
  \centering
  \includegraphics[width=0.9\textwidth]{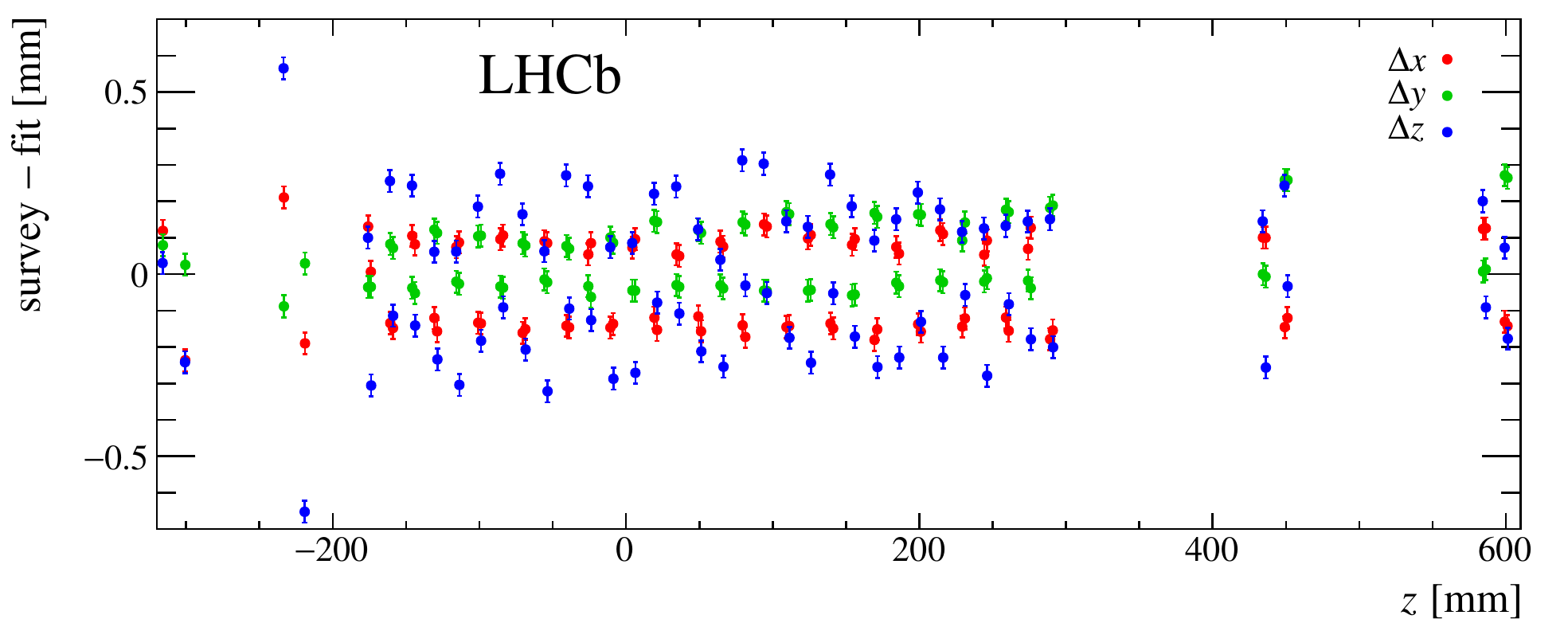}
  \caption{
  Differences in the sensor locations relative to the survey specifications. The observed deviations are known from VELO alignment studies and accounted for during reconstruction.
  }
  \label{fig:mods}
\end{figure}

Figure~\ref{fig:mods} shows the differences in the $x$, $y$, and $z$ locations of each module with respect to the VELO survey specifications.
The largest discrepancies observed in $z$ are $\approx 0.6\mm$ in two of the pile-up sensors~\cite{ref:pileup}, while all standard sensors are consistent with their survey positions to $\lesssim 0.3\mm$~\cite{ref:veloalign}.
The $y$ positions are all found to be consistent with the survey values to $\approx 0.1\mm$; however,
one can see that the VELO is not fully closed in $x$~\cite{ref:velo}.
These observed deviations from the survey specifications are known from the software alignment procedure and taken into account during the reconstruction, and reproducing them provides validation of the SV-based approach.

\begin{figure}[t]
  \centering
  \includegraphics[width=0.9\textwidth]{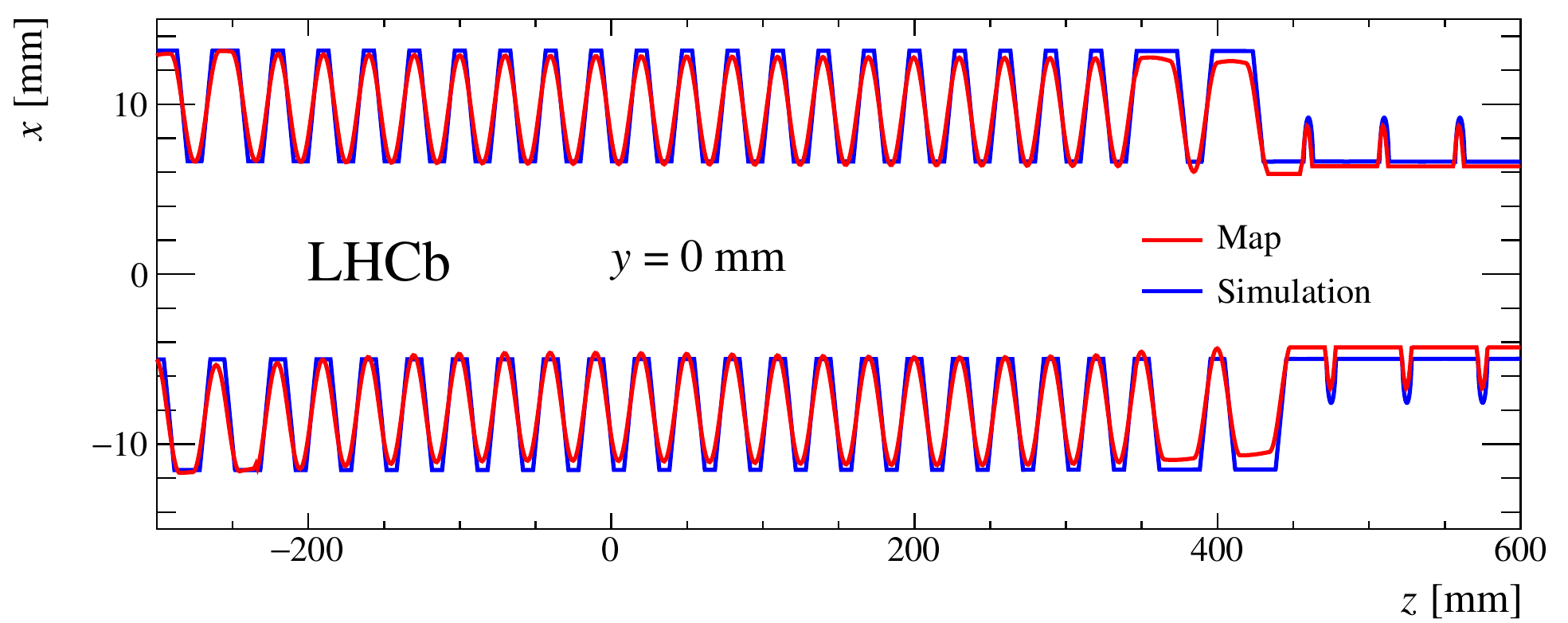}
  \caption{
  Comparison between the RF-foil map at $y=0\mm$ and the description in LHCb simulation.
  }
  \label{fig:foil}
\end{figure}

The shape of the RF-foil in the $xy$-plane is roughly a semi-circle about the origin that transitions into straight lines that extend out away from the origin at fixed $x$ values.
The parametrization employed here describes these transitions using additional semi-circles (interpolation using bicubic splines was also tried, but found to provide a worse description of the data).
The $xy$ distributions of SVs are fit in 1\mm wide slices in $z$, where SVs consistent with originating from a module are removed, and in each of the 1066 slices four parameters are determined.
As with the modules, the thickness of the RF-foil is ignored when building the map, but accounted for as an uncertainty in the true origin point when determining the material-interaction $p$-value.
The $z$ dependence of each RF-foil parameter is then fit to the following empirical functions:
the $y$ locations of all semi-circles are found to be well described by simple linear functions,
while the remaining parameters are fit to a combination of sinusoidal and polynomial functions, except in the forward-most region, {\em i.e.}\ at large $z$, where dedicated functions are needed to capture additional structure in the RF-foil.
Figure~\ref{fig:foil} shows an example comparison of the RF-foil map created here to the description of the RF-foil in the LHCb simulation.
The simulation clearly uses a simplified shape---but also one that fails to describe the sizable deviations from the design shape in the forward-most region.

\begin{figure}[t]
  \centering
  \includegraphics[width=0.99\textwidth]{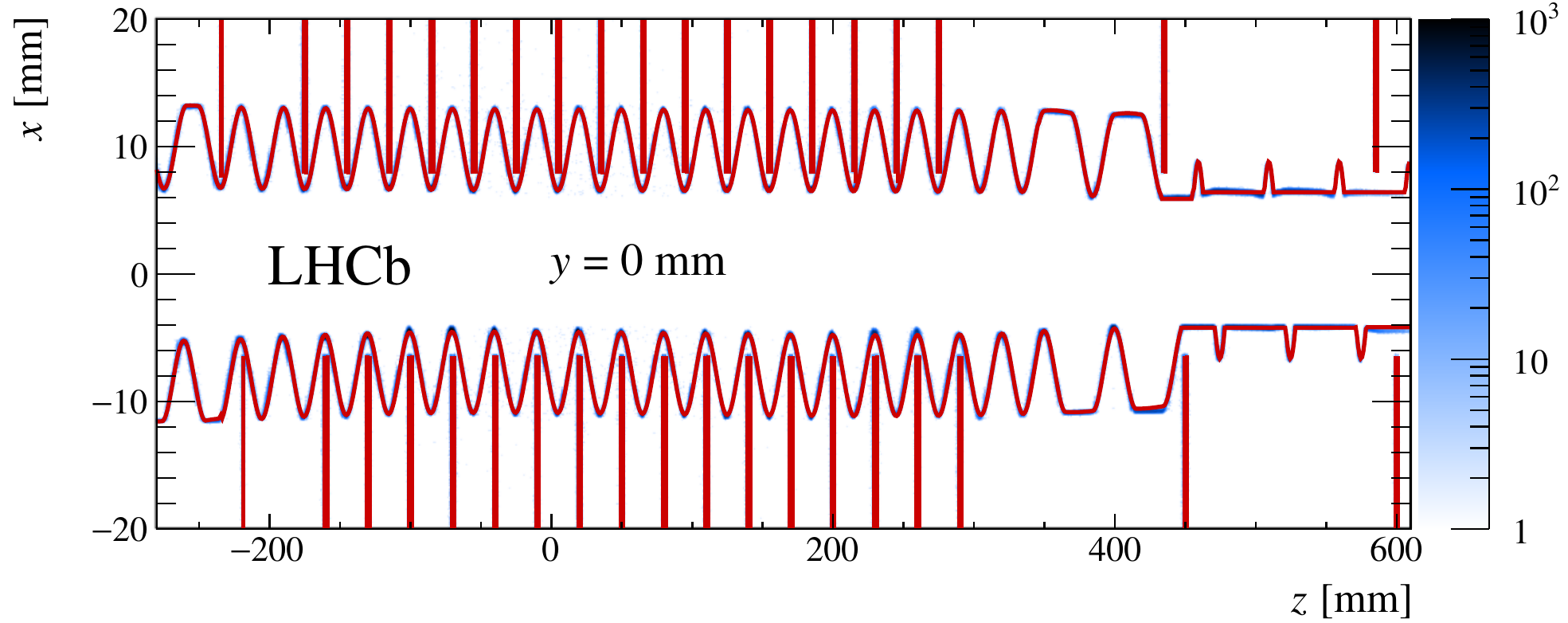}
  \includegraphics[width=0.99\textwidth]{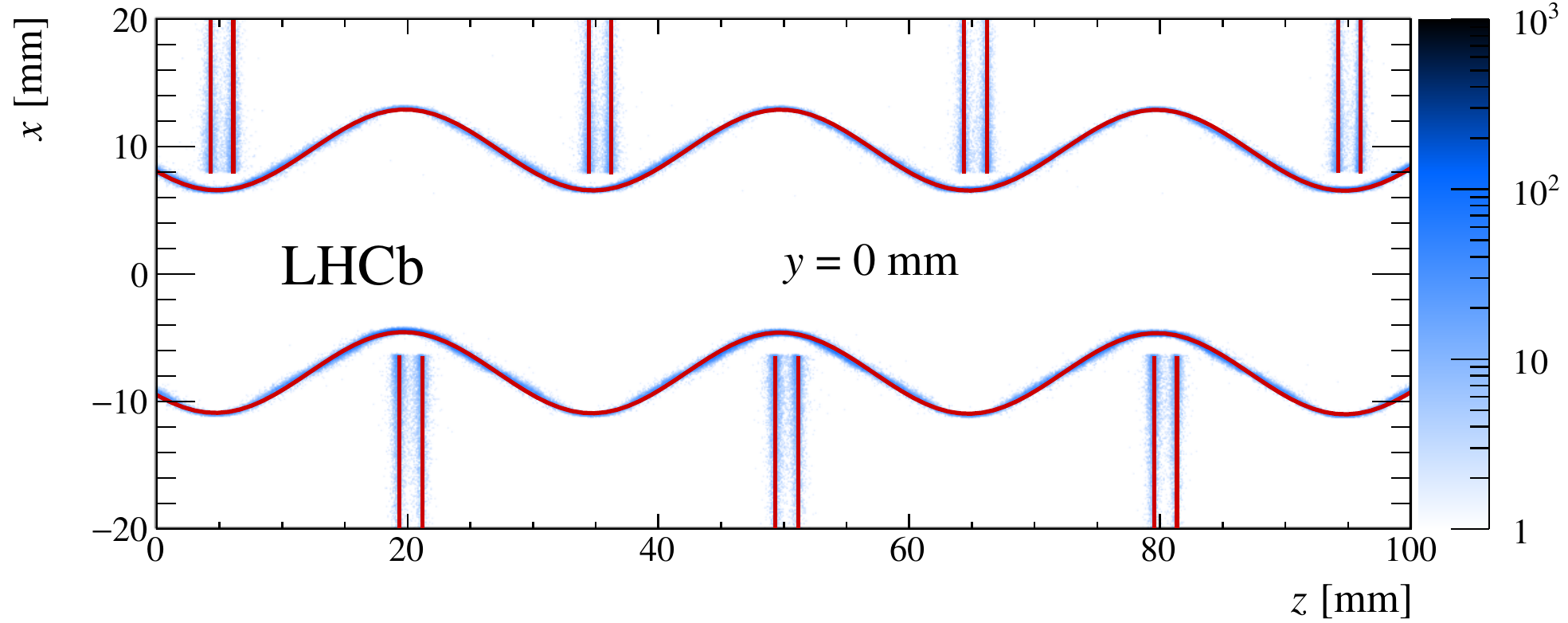}
  \includegraphics[width=0.48\textwidth]{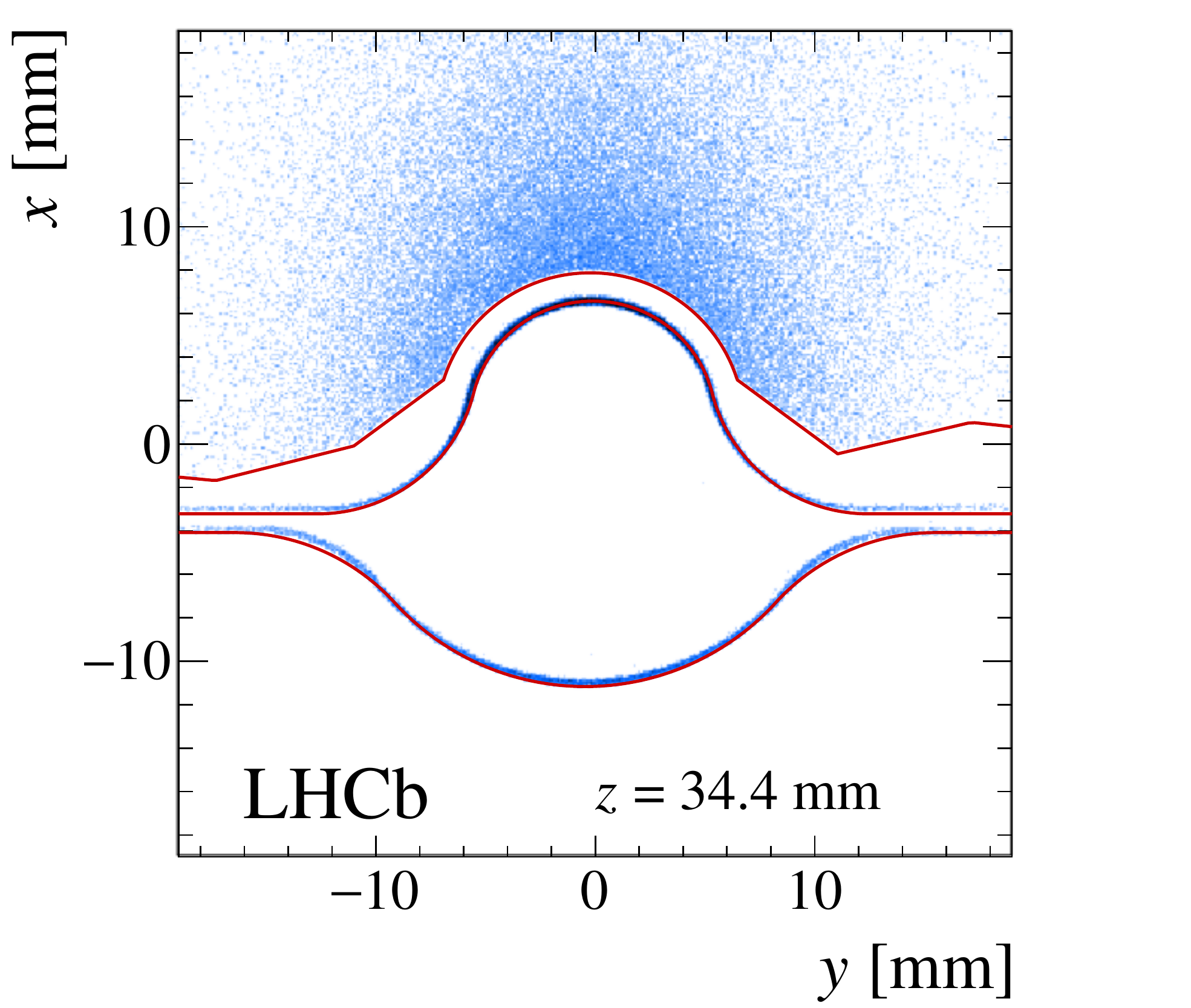}
  \includegraphics[width=0.48\textwidth]{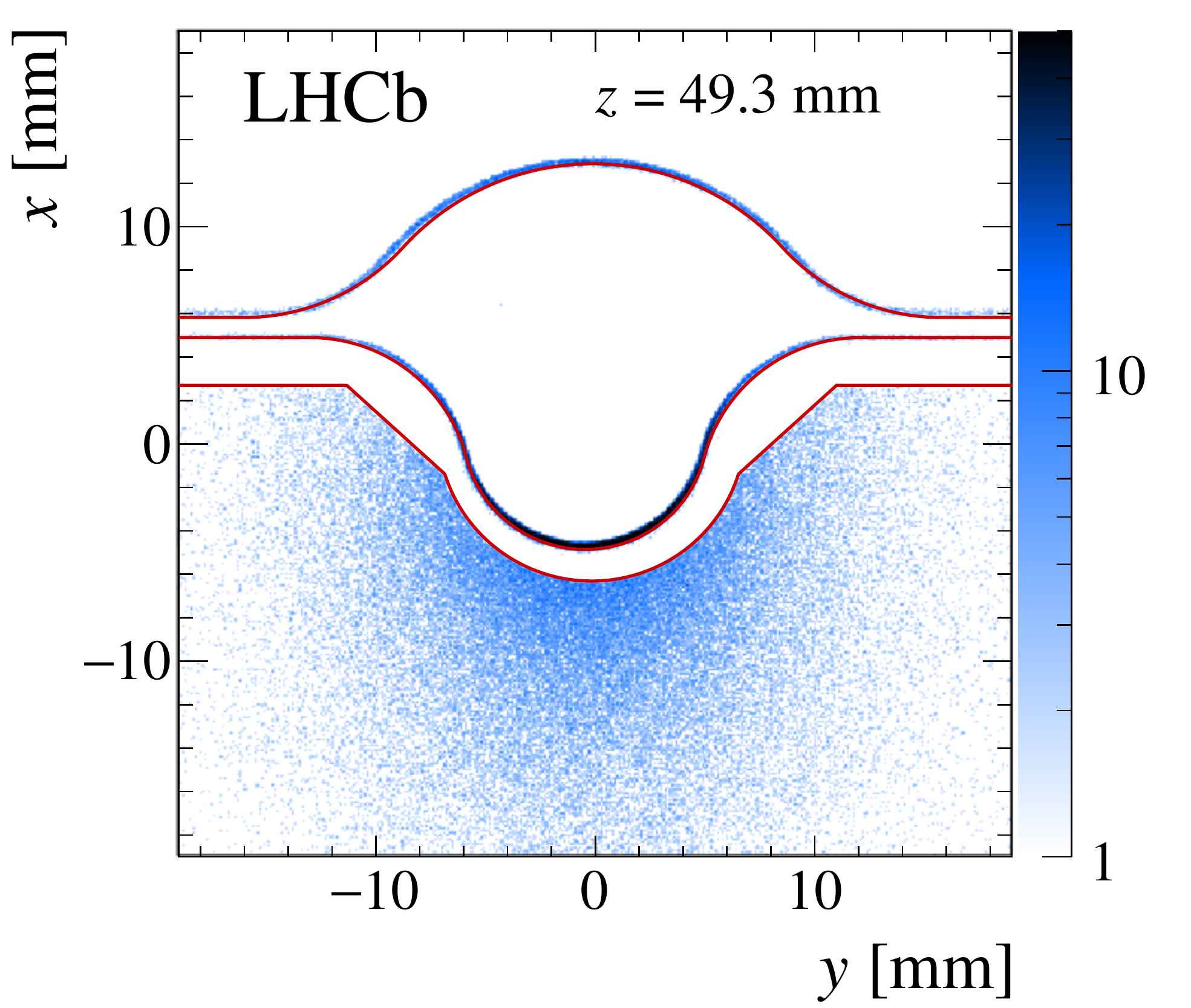}
  \caption{
  Example comparisons between the (filled bins) reconstructed SV locations and (red lines) the material map for:
  (top) $x$ versus $z$ near $y=0\mm$;
  (middle) same as the top but zoomed in on the ${0 < z < 100\mm}$ region;
  (bottom left) $x$ versus $y$ near the left-half $r$ sensor at $z=34.4\mm$; and
  (bottom right) $x$ versus $y$ near the right-half $\phi$ sensor at $z=49.3\mm$.
  For visual clarity:
  in the top panel, the red vertical lines denote the center of each sensor, while in the bottom panels only the edges of the sensors are shown;
  and
  in all panels, only the nominal RF-foil position is shown; {\em i.e.}\ the red RF-foil curves do not display its thickness.
  }
  \label{fig:map}
\end{figure}

Figure~\ref{fig:map} shows some example comparisons of the material map to the observed SV distributions.
Overall, the map describes the data well.
There are some small discrepancies in the forward-most region, where some features of the RF-foil are not fully described; however, high precision is not required in this region, since SVs reconstructed here are less precisely determined due to the large separation between the modules at large $z$.
Finally, given that the RF-foil is $\approx 0.3\mm$ thick, that its locations are determined in fits that integrate over $z$ regions that are 1\mm wide, and that its shape varies along the length of the VELO in ways that are not all fully captured by our parametrization at all $z$ values, an uncertainty of 0.5\mm is assigned to the RF-foil location in the $xy$ plane when vetoing material interactions.

\newpage

\section{Material Probability}
\label{sec:prob}

In this section, a method is presented that reports a $p$-value for the hypothesis that the true SV location is consistent with a point in space occupied by VELO material.
The inputs used to determine this $p$-value are the 3-D location of the SV and its uncertainty, and the 3-D material map with the appropriate adjustments for the location of the detector during each fill.
Several possible metrics were considered, with the best performance achieved using the harmonic mean of uncertainty-weighted material distances
\begin{equation}
D = 6 \left[\sum\limits_{i=1}^6 D_i^{-1}\right]^{-1} \!\!\!\!\!,
\end{equation}
where each $D_i$ is defined as
\begin{equation}
  D_i = {\rm min}\left[\sqrt{\left(\frac{x_{\rm m}^i-x_{\rm sv}}{\sigma_x}\right)^2 +  \left(\frac{y_{\rm m}^i-y_{\rm sv}}{\sigma_y}\right)^2 + \left(\frac{z_{\rm m}^i-z_{\rm sv}}{\sigma_z}\right)^2 }\,\right],
\end{equation}
for a VELO material element  described by $\{\vec{x}_{\rm m}\}$, given an SV location $\vec{x}_{\rm sv}$, and  uncertainties in each direction $\vec{\sigma}$, which include contributions from both the material and SV locations.
The six material elements considered are the sensors in each VELO half (two sensors), along with the RF-foil in each half where the positive and negative $z$ directions from the SV $z$ location are considered separately (four RF-foil elements).
Each sensor is represented by its design shape in the $xy$ plane and a $z$ location with zero thickness.
The minimum in the above equation considers the expression in square brackets for every point on the sensor, taking the smallest value.
For the RF-foil, the minima are found in a conceptually similar way, but 3-D numerical searches are required to find them due to the complicated shape of the RF-foil.

The expected $D$ distribution can be obtained for any data sample as follows: reconstructed SVs in the data sample that are close to material can be repeatedly resampled, taking the closest material point as the true origin and sampling the SV locations according to their uncertainties.
Non-Gaussian effects are accounted for by taking the resampling distributions from simulation, {\em i.e.}\ the differences in the true and reconstructed SV locations are not assumed to follow Gaussian distributions with widths given by the SV resolution. Instead these distributions are obtained from the full LHCb simulation.\footnote{The SV-fit $\chi^2$ values are required to be good, which highly suppresses the non-Gaussian contributions and mitigates any potential mismodelling of these effects in the simulation. Additionally, non-Gaussian scattering effects are suppressed even further in analyses that apply criteria to the consistency of the decay topology.}
This data-driven approach must be applied for each data sample, since the $D$ distribution depends strongly on the SV resolution, which is correlated with the decay opening angle, number of decay products, and other decay-specific features.

\begin{figure}[t]
  \centering
  \includegraphics[width=0.6\textwidth]{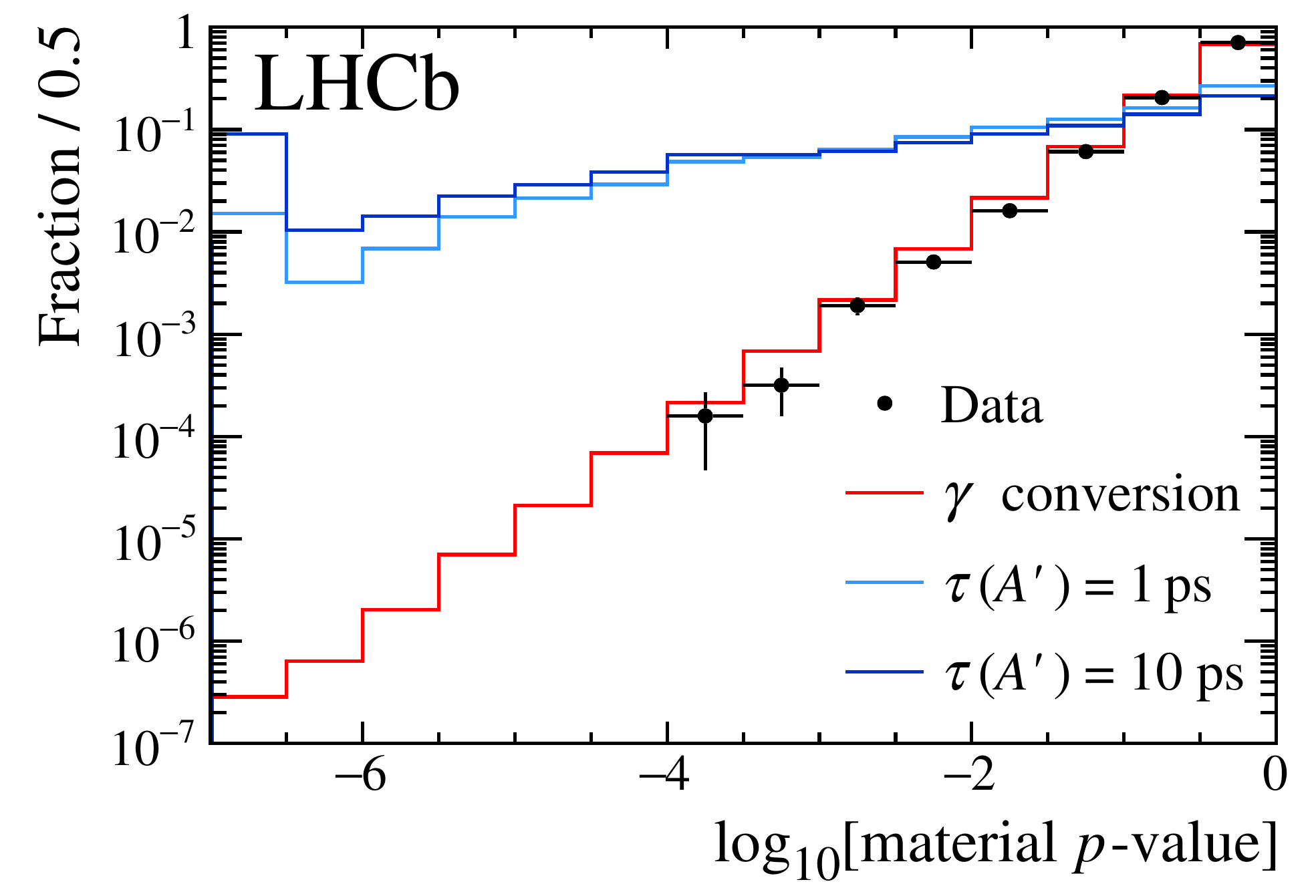}
  \caption{
  The normalized photon conversion $p$-value distribution obtained using a subsample of data from an LHCb long-lived dark-photon search~\cite{ref:lhcbatomm}.
  The data are consistent with the photon-conversion hypothesis.
  Some example dark-photon distributions are also shown for lifetimes of 1 and 10\,ps, showing good separation between potential exotic signals and photon conversions.
  }
  \label{fig:pval}
\end{figure}

Figure~\ref{fig:pval} shows a validation of this procedure from a search for long-lived dark photons decaying to dimuon final states~\cite{ref:lhcbatomm}.
The $A^{\prime}\to\mu^+\mu^-$ candidates are built from muons that are inconsistent with originating from the $pp$ collision, satisfy stringent muon-identification criteria, and have ${\pt(\mu)>0.5\gev}$ and $p{(\mu)>10\gev}$.
To avoid potential bias in the SV locations and their uncertainties, the following criteria are applied:
the first two hits on each muon track are required to be in an $r$--$\phi$ sensor pair,
the first hit on each track is required to be in the first VELO sensor intersected by its trajectory,
the first hits on the two muons must be in the same sensor module,
the two muons cannot share more than 8 hits in the VELO to remove clones,
and the SV is required to be upstream of the first hits on each muon track.
The subsample shown in Fig.~\ref{fig:pval} uses only candidates with an SV location at least 5\mm from the beam line and with $m(\mu^+\mu^-) < 0.25\gev$ to suppress heavy-flavor and double-misidentified $K_S\to\pi^+\pi^-$ backgrounds.
The data are consistent with the photon-conversion hypothesis.
The search presented in Ref.~\cite{ref:lhcbatomm} applied a mass-dependent criterion on $D$ that reduced the contribution from conversions to a negligible level, while maintaining good signal efficiency.
This procedure makes it possible to perform nearly background-free searches for many proposed types of long-lived exotic particles, greatly enhancing the physics discovery potential {\em c.f.}\ simply removing SVs in predefined material regions.

\section{Summary}
\label{sec:sum}

In summary, a study of the LHCb VELO material based on secondary hadronic interactions was presented, and a high-precision map of the VELO material was built.
The analysis used secondary interactions of hadrons produced in beam-gas collisions collected during special run periods where helium or neon gas was injected into the beam-crossing region.
Material interactions occur along the entire length of the VELO in such events, rather than just near the $pp$-interaction region.
Using this  material map, along with properties of a reconstructed SV and its constituent tracks, a $p$-value can be assigned to the hypothesis that the SV originates from a material interaction.
This approach was recently used to veto photon conversions to $\mu^+\mu^-$ in a search for dark photons at LHCb~\cite{ref:lhcbatomm}.
The procedure makes it possible to perform nearly background-free searches for many proposed types of long-lived exotic particles.

\acknowledgments

A special acknowledgement goes to
all of our LHCb collaborators who contributed to obtaining the results presented in this paper.
Specifically, we thank Plamen Hopchev for help developing the special reconstruction used in this study;
and
Yuri Guz,
Christian Joram,
and
Rosen Matev
for providing useful feedback on this article.
We express our gratitude to our colleagues in the CERN accelerator departments for the excellent performance of the LHC.
We thank the technical and administrative staff at the LHCb institutes.
We also acknowledge support for the LHCb collaboration from CERN and from the national
agencies: CAPES, CNPq, FAPERJ and FINEP (Brazil); MOST and NSFC
(China); CNRS/IN2P3 (France); BMBF, DFG and MPG (Germany); INFN
(Italy); NWO (The Netherlands); MNiSW and NCN (Poland); MEN/IFA
(Romania); MinES and FASO (Russia); MinECo (Spain); SNSF and SER
(Switzerland); NASU (Ukraine); STFC (United Kingdom); NSF (USA).  We
acknowledge the computing resources that are provided by CERN, IN2P3
(France), KIT and DESY (Germany), INFN (Italy), SURF (The
Netherlands), PIC (Spain), GridPP (United Kingdom), RRCKI and Yandex
LLC (Russia), CSCS (Switzerland), IFIN-HH (Romania), CBPF (Brazil),
PL-GRID (Poland) and OSC (USA).
Individual LHCb groups or members have received support from AvH Foundation (Germany), EPLANET, Marie Sk\l{}odowska-Curie Actions and ERC (European Union), ANR, Labex P2IO, ENIGMASS and OCEVU, and R\'{e}gion Auvergne-Rh\^{o}ne-Alpes (France), RFBR and Yandex LLC (Russia), GVA, XuntaGal and GENCAT (Spain), Herchel Smith Fund, the Royal Society, the English-Speaking Union and the Leverhulme Trust (United Kingdom).


\begin{thebibliography}{99}
  \bibitem{ref:velo} LHCb VELO group, {\em Performance of the LHCb Vertex Locator}, JINST {\bf 9} (2014) P09007. [arxiv:1405.7808]
  \bibitem{ref:atomm} P. Ilten, Y. Soreq, J. Thaler, M. Williams, and W. Xue, {\em Proposed inclusive dark photon search at LHCb}, Phys.\ Rev.\ Lett.\ {\bf 116}, 251803 (2016). [arxiv:1603.08926]
  \bibitem{ref:atoee} P. Ilten, J. Thaler, M. Williams, and W. Xue, {\em Dark photons from charm mesons at LHCb}, Phys.\ Rev.\ D {\bf 92} (2015) 115017. [arxiv:1509.06765]
  \bibitem{ref:bcjets} LHCb collaboration, {\em Identification of beauty and charm quark jets at LHCb}, 	JINST {\bf 10} (2015) P06013. [arxiv:1504.07670]
  \bibitem{ref:veloalign} S. Borghi {\em et al}, {\em First spatial alignment of the LHCb VELO and analysis of beam absorber collision data}, Nucl.\ Instrum.\ Meth. {\bf A618} (2010) 108-120.
  \bibitem{ref:veloap} M. Ferro-Luzzi, T. Latham, and C. Wallace, {\em Determination of the aperture of the LHCb VELO RF foil}, LHCb-PUB-2014-012.
  \bibitem{ref:atlasmat} ATLAS collaboration, {\em A study of the material in the ATLAS inner detector using secondary hadronic interactions}, JINST {\bf 7} (2012) P01013 [arxiv:1110.6191]; {\em A measurement of material in the ATLAS tracker using secondary hadronic interactions in 7 TeV $pp$ collisions}, JINST {\bf 11} (2016) P11020. [arxiv:1609.04305]; {\em Study of the material of the ATLAS inner detector for Run 2 of the LHC}, JINST {\bf 12} (2017) P12009. [1707.02826]
  \bibitem{ref:lhcbatomm} LHCb collaboration, {\em Search for dark photons in 13 TeV $pp$ collisions}, Phys.\ Rev.\ Lett.\ {\bf 120} (2018) 061801. [arxiv:1710.02867]
  \bibitem{ref:atlasll} ATLAS collaboration, {\em Search for massive, long-lived particles using multitrack displaced vertices or displaced lepton pairs in $pp$ collisions at $\sqrt{s} =8$\,TeV with the ATLAS detector}, Phys.\ Rev.\ {\bf D92} (2015) 072004. [1504.05162]
  \bibitem{ref:smog2} C. Barschel, {\em Precision luminosity measurements at LHCb with beam-gas imaging}, PhD thesis, RWTH Aachen University, 2014, CERN-THESIS-2013-301.
  \bibitem{ref:lhcb} LHCb collaboration, {\em The LHCb detector at the LHC}, JINST {\bf 3} (2008) S08005; {\em LHCb detector performance}, Int.\ J.\ Mod.\ Phys.\ {\bf A30} (2015) 1530022. [arXiv:1412.6352]
  \bibitem{ref:smog1} M. Ferro-Luzzi, {\em Proposal for an absolute luminosity determination in colliding beam experiments using vertex detection of beam-gas interactions}, Nucl.\ Instrum.\ Meth.\ {\bf A553} (2005) 388.
  \bibitem{ref:lumi} LHCb collaboration, {\em Precision luminosity measurements at LHCb}, JINST {\bf 9} (2014) P12005. [arXiv:1410.0149]
  \bibitem{ref:pileup} S. Oggero, {\em Beauty in the crowd: Commissioning of the LHCb pile-up detector and first evidence of $B_s^0 \to \mu^+\mu^-$}, PhD thesis, Vrije Universiteit Amsterdam, 2013, CERN-THESIS-2013-225.
\end{thebibliography}
\end{document}